# A LEGAL PERSPECTIVE OF E-BUSINESSES AND E-MARKETING FOR SMALL AND MEDIUM ENTERPRISES (SMEs)

Muneeb Iqbal[1], Atif Ali Khan[2], Oumair Naseer[3]

[1]The School of Architecture, Computing and Engineering University of East London, UK
muneeb.iqbal@live.co.uk
[2]School of Engineering, University of Warwick, Coventry, UK
Atif.Khan@warwick.ac.uk
[3]Department of Computer Science, University of Warwick, Coventry, UK
o.naseer@warwick.ac.uk

## ABSTRACT

*Electronic businesses are witnessing enormous growth as more and more people are switching to online platforms. The widespread use of Internet has opened new channels to operate trade for many businesses. Also electronic marketing has become a proven channel of passing on the word to the customers. Legal and ethical issues quickly become an area of concern. In this research recommendations are made to harmonize IT and Internet Laws. A novel approach is proposed to promote legal risk management culture in organizations. It begins with revising current state of regulations surrounding eBusinesses and electronic marketing. The proposed approach offers risk management by considering risk mitigation strategy, educating people and use of information technology. Monitoring compliance requirements are met by reviewing the latest changes in regulations and rewarding the employees who ensures the successful implementation of the strategy.*

## 1. INTRODUCTION

The process of Risk Management is broken down into three parts: i) identification of risk sources, ii) assessment of their effects (risk analysis), iii) development of management response to risk (Perry, 1986). The fundamentals of risk management are the same but their implementations are being adapted according to the needs of current era's risks such as legal risks. eBusinesses and eMarketing campaigns quickly become vulnerable to many legal threats. As soon as an SME's online business starts to grow it finds itself surrounded by many regulations and statutory compliance requirements. There are laws that govern the business industry. Also, there are regulatory bodies for the implementation of rules. Some regulations are common to all businesses and some are unique to online businesses. Due to the increase in the customer's requirements, many companies heavily invest in eMarketing. There are many regulations that eMarketing industry has to follow. A regulatory body named "Advertising Standards Authority" governs the laws of Marketing. It is vital for an eBusiness to understand the legal threats that business may have and develop strategies to avoid and eliminate legal risks.

In the first section of this research paper, the importance of legal and ethical aspects of businesses; especially electronic businesses and electronic marketing are discussed. Some of the regulations surrounding eBusiness and eMarketing are highlighted in the next section. Later a statistical review is provided to show complaints arising due to non-compliant advertising through several advertisement media. Some of the common law breakings are named which are solicited and unsolicited such as; commercial communications, information to be provided about placing of orders by electronic means, information to be provided about where contracts are concluded by electronic means etc (The Electronic Commerce (EC Directive) Regulations 2002).

The primary goal of this research is to study legal risk management for eBusinesses and eMarketing strategies. To achieve the aim following objectives are set:

    i.     Choose a risk management model.
    ii.    Understand and evaluate legal risks.
    iii.   Carry out legal risk analysis.
    iv.   Study the current state of regulations surrounding eBusinesses and eMarketing.
    v.    Propose an approach to promote legal risk management in an organization.
    vi.   To give recommendations to improve eBusiness Legal Framework.
    vii.  To give recommendations for SMEs to run eBusinesses legally.

## 2. LEGAL RISK MANAGEMENT

Risk Management can be applied on any kind of operation. Legal Risk Management is growing very rapidly. These risks are not self-contained. There may be other legal aspects to commercial and technical risks. The risks which are basically legal, may also be combined with technical and commercial elements (Burnett, 2005). A 5-process legal risk management model adapted by the Department of Justice and Attorney-General, Queensland Government for Safety Code of Practice Risk Management, 2010 is as follows:

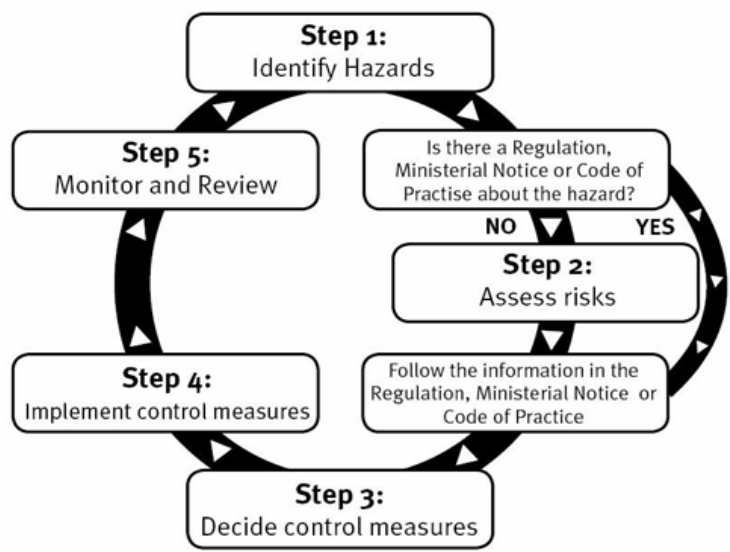

Figure 1: 5-Process Legal Risk Management Model adapted by Department of Justice and Attorney-General, Queensland Government for Safety Code of Practice Risk Management.

# 3. LEGAL RISKS: IDENTIFICATION AND ANALYSIS

Some of the major identified risks are as follows:

**3.1. Selling Compliance:** There are laws that govern sales and purchases by any means. In the UK products and services sales over the Internet are protected by many legislations and regulations in the public interest. Non-compliance in delivering product and services can quickly become a legal risk for the business. and is highly vulnerable to cause damage in the shape of financial damage, loss of customer trust or loss of reputation in the market.

**3.2. Intellectual Property:** Intellectual Property (IP) is an issue of concern for centuries. There are four main types of IP i.e. Copyright, Patents, Designs and Trade Marks (IPO, 2005). With the advancements of computing technologies, software development and Internet, requirements for IP compliance are a vital aspect of the legal strategy for a company (Hanel, 2006). According to Hayes (2001) there are two main dimensions of discussion in the subject of Intellectual Property. Firstly an eBusiness must ensure compliance to IP rules and regulations and secondly a set of procedure should be in place to protect and police business's Intellectual Property.

**3.3. Electronic Marketing:** Marketing is an operation, which is often considered as the backbone of a trade. It assists in brand awareness, revenue generation and client base generation. There are obligations on businesses that should be met for legal marketing. It is a legal risk that is most sensitive because due to brand image potential customer perceives product being advertised. Search Engine Marketing is rather proactive as it targets customer that are currently looking for the product and may complete the purchase at the same point. However all eMarketing strategies should comply legally and be in accordance of natural consumer behaviour.

**3.4. Information Security:** A key element to stay competitively ahead of others is to ensure security, integrity and safety of business communication and customer information. A fundamental principle is to ensure that any communication sent must get to its recipient unaltered. It is equally important for businesses as well as individuals to have the surety that the origin of information or communication is verifiable, secrecy of its contents, and to have confidence in its accuracy. Information security is also important to gain competitive advantage ethically and to comply with legal requirements.

**3.5. Data Protection:** eBusinesses which stores and processes consumer data must comply with the Data Protection Act 1998 and register with The Information Commissioner's Office (ICO, 2011). It is a legal risk for a business to ignore the legal requirements for data protection. As eBusinesses are generally global in nature and processes customer's data from outside the UK as well, optimum level compliance should be kept.

# 4. Current State of Regulations surrounding EBusinesses and EMarketing

Some of the major legal risks identified to eBusinesses and eMarketing along with their current state of regulations are as follows:

## 4.1. Selling Compliance

eBusinesses are governed by legislations such as The Electronic Commerce (EC Directive) Regulations 2002 and The Consumer Protection (Distance-Selling) Regulations 2000 which provides a comprehensive guide on selling online legally. Customer should be provided accurate information about the merchant and the product attributes before the purchase is completed. The minimum information that needs to be provided legally is: identity of the supplier, payment requirements, description of the main characteristics of the product or services and right of cancellation (The Consumer Protection (Distance Selling) Regulations 2000, s(7). In modern selling models the concept of up selling and product add-on is common and many online retailers try to sell additional products, services or paid memberships to their customers. Office of Fair Trading takes such sales very seriously. Consumers must have clear information provided about any charges or subsequent contract they may enter into while purchasing a product online. According to OFT (2005) "The companies have, without accepting that they were in breach of the DSRs, given undertakings that they:
Will not charge a sum to a customer's credit or debit card without their express authority and the full knowledge of the purpose, amount and timing of the payments without having reasonable cause to believe that they have a right to payment and will not promote similar schemes, which fail to give customers full details of the scheme including that it is a negative option marketing scheme and that customers have to choose to opt out. If any of the parties breach the undertakings the OFT could seek an injunction against them in the High Court. Failure to obey an injunction could result in proceedings for contempt of court." Therefore business manager should be careful when making strategies for increased online sales else they could face penalties by the courts resulting various damages as described above.

In the selling process another important aspect is 'commercial communication' before a sale contract has concluded between the two parties i.e online merchant and the customer. Section 7 of The Electronic Commerce (EC Directive) Regulations 2002 provides guidelines for commercial communication to make the contract lawful. The communication must be clearly identifiable and unambiguous. eBusinesses are heavily based on automated responses generated that follow a business work flow. The Customer Relationship Management System (CRM) should not only assist in efficient management of customer information, complete sales confirmation process, relationships between customers and suppliers etc. but also legal obligations of workflow processes must be met.

## 4.2. Intellectual Property, Copyrights, Trade Marks and Patents

Copyright, Designs and Patents Act 1988, and the Copyright and Related Rights Regulations 2003 provide extensive guidance for Copyright regulations in UK. According to Copyright, Designs and Patents Act 1988, Copyright is property right, which protects original literacy, dramatic, musical, or artistic work through any

mediums. Some of the issues regarding eBusiness copyright that may arise are disputes about the work ownership between employee and employer, web content provision, copyright infringement etc. Ownership of the work is with the author, creator, designer or developer of the work unless the author is an employee of a company. In such case the ownership of the work is with the company unless otherwise agreed in the employment contract. For an eBusiness the company should clearly state the ownership in the employment contract to avoid any possible dispute that may arise in future. Similarly if the employee has done a copyright infringement the company is liable for the damages. eBusiness should place a check and balance procedure to ensure that the staff is adhering to the copyright laws. Awareness could be increased through extensive training provided, organising IP awareness conferences and promote a law-abiding culture in the organisation.

The global nature of Internet has widespread copyright infringement. eBusinesses often suffer from copyright infringements. A policy should be developed and implemented to protect copyright (Yoon, 2001). Watermarking is useful in some types of work. Through research and development various technologies can be identified or developed to police copyright infringement. Many software are available offering copy protect. (Trade Mark Act, 1994) To stay protected from this risk an eBusiness should aim to protect its vital data such as: media documents, training materials, e-learning contents, trade and business documents, company presentations, creative files and project files. Hence, copyright is a vital aspect for an eBusiness and can quickly become a prominent risk, if certain measures are not taken to protect all dimensions of its violations.

Brand profiling can play important role in the success of any business. Similarly Internet brand management is the key in marketing strategy for an eBusiness. Violations in Trade Marks are a threat to brand protection on the Internet. Trade Marks Act 1994 provides extensive guidelines to benefit from Trade Marks registration and the rights protection gained after registration. According to the Trade Mark Act 1994, "a registered trade mark is a property right obtained by the registration of the trade mark. Under this Act and the proprietor of a registered trade mark has the rights and remedies provided by this Act". For an eBusiness, the major risks are; to protect their trademarks and avoid violations, to effectively manage trademark registration, to avoid registration application refusals, renewals and alterations. There are also costs associated to such right protections. Information Technology have revolutionised almost every aspect of our lives. Trademarks can be efficiently managed by the use of Trademark Management Systems. There are many available of-the-shelf in the market as well as many companies develop their own according to the nature of their work. WebTMS and CPi are just two examples of such systems that can help assist in better management of company's Trademarks. However, the choice of recommended processes to avoid Trademarks infringement risk depends on the nature of the trademarks and the legal risk management strategy.

To stay competitive in the industry many companies invest in Research and Development of new innovations and inventions. Patents Act 2004 protects the inventions through legal protection. We have witnessed vigorous advancements in the Information Technology. New technologies emerge fairly quickly. Patents should be legally protected to keep the competitive advantage. It quickly becomes a risk to the technological advancements of an eBusiness. On the other hand, there are huge costs involved in providing legal protection to the inventions, which is often a burden on SMEs, This risk is often ignored. An eBusiness, which is proactive in R&D but yet to earn enough market shares to be able to fund R&D and protect inventions legally, this could lead it to financial misbalance and legal complications risk. Although Patent protection exists in the society to encourage inventions, however, approach to it is limited to rather large organisations. Only a few SMEs make use of such protections to encourage inventions.

In UK, Intellectual Property Office is responsible to ensure Patent related laws. Extensive information, guide and support are provided by IPO to those who wish to patent their inventions. An interesting point is that Patents protected in the UK by IPO are only protected in UK and not outside the jurisdiction. In cases, this needs to be protected within the UK as well as abroad; it is up to the management to decide whether their patent needs to be protected within the UK only or it needs overseas protection. This heavily depends on the fact that if the business intends to provide its goods or services in the target country. eBusinesses are usually global in nature and need cross border protection. It is often a vulnerable risk to unprotected inventions or potential products or services that may arise due to research and development within an organisation.

The European Patent Organisation is an intergovernmental organisation that was set up on 7 October 1977 on the basis of the European Patent Convention (EPC) signed in Munich in 1973. It has two bodies, the European Patent Office and the Administrative Council, which supervises the Office's activities (EPO, 2011). There 38 member states that recognise and legally protect patents accepted by EPO. For online businesses that supply goods or services cross border, unprotected patents and locally protected patents stays a threat as others may violate. On the other hand, eBusinesses should respect others copyright, trademarks and patents to avoid any criminal activities. Information Systems for patents management can assist in effective management of patent registration, updating and renewals.

### 4.3. Electronic Marketing

According to a research carried out by Office of National Statistics (2011), 41.26 million adults who had ever used the Internet by the second quarter of 2011, representing 82.3 percent of the adult's population. From these finding it is obvious that a large population is using Internet therefore electronic marketing has become a massive medium for advertising. Legal compliance steps in this operation of a business too making it another obvious legal risk for a business if compliance is ignored. Until recently the policing of marketing code of practice did not include online media. In 2011 the remit of the ASA was extended significantly to deliver more comprehensive consumer protection online (ASA, 2011).

Some of the legal risks attached to marketing materials are misleading advertisements, social responsibility and the protection of children. It is ensured by ASA that same high standards are met for advertisements on own websites, marketing communications in other non-paid-for space under the advertiser's control, such as social networking sites like Facebook and Twitter and marketing communications on all UK websites, regardless of sector, type of businesses or size of organisation (ASA, 2011).

Two the most successful marketing categories are direct marketing and multi-level marketing. In the electronic marketing those two strategies are widely used. Both of those have legal risks attached with them. For direct marketing consumer data is needed. The use of personal information regarding marketing has laws governing purpose of use of information by Data Protection Act 1998, interference into individual's privacy by The Privacy and Electronic Communications (EC Directive) Regulations 2003. Section 22, 23, and 24 provides guides for legal requirements when using emails for marketing purposes. It could rapidly cost large amounts of fines and other penalties to a company if non-compliant marketing activities are carried out, therefore marketing strategy should be in line with legal requirements to avoid certain damages.

### 4.4. Information Security
The unauthorized alteration of contents of communication, malicious communication on behalf of an organization etc. are some of the risks attached to information security and cause reputation damage, financial losses, loss of trust and loss of information to the company. There are legal obligations in order to protect data. Data Protection Act 1998 protects personal information of pubic.

Electronic signatures have become legally recognized in many jurisdictions. With the expansion of online business communications many contracts are completed electronically. eBusinesses should ensure that contracts are legally binding. Throughout EU eSignatures are harmonized legally through the EC Electronic Signatures Directive (1999/93/EC). The directive is implemented in UK through The Electronic Communications Act 2000 and the subsequent Electronic Signatures Regulations 2002. Unless the legal requirements are met the contracts completed through eSignatures may be void and become a threat for the business.

### 4.5. Data Protection
Data protection compliance is very vital in almost all sorts of businesses including eBusinesses. Data Protection Act 1998 protects consumer data and provides eight most important principles of data protection. However in EU a harmony of data protection exists through Directive 95/46/EC on data protection (the "Data Protection Directive"), Directive 2002/58/ EC on privacy and electronic communications ("the ePrivacy Directive"), and Directive 2009/136/EC (Jones and Tahiri, 2011). Iceland, Liechtenstein and Norway have also agreed to bind by the Data Protection Directive. Recently rules for the use of cookies by the websites have been put into practice. Until recently the general perception of rejecting cookies was to manually set the browser to reject cookies by the visitor. This is an 'opt-out' case; however the ePrivacy Amendment Directive turns it into 'opt-in' option where the user should be informed about the use of cookies upfront and provide the option to reject it if the users wished not to download cookies (Jones and Tahri, 2011). Websites that are currently using cookies should adapt to the new requirements to avoid penalties and non-compliance. It should be ensured that following operations of data processing are met compliant in line with the legal requirements:

i. Fair and lawful processing of data.
   ii. Collection of data for specific lawful purposes and no further processing that is incompatible with the original purpose.
   iii. Maintaining adequacy and relevancy of personal data collection and processing.
   iv. Maintaining accuracy.
   v. Not holding the data for longer that what it is actually needed for.
   vi. Appropriate measures are taken to secure the data.
   vii. No transferring of data outside EEA unless that country ensures adequate level of protection and freedom of data.

## 5. AN APPROACH TO PROMOTE LEGAL RISK

In the previous sections some of the major legal risks for eBusinesses are identified and assessed. The current legal framework is discussed and important aspects that should be legally complied are pointed out. In this section an approach is proposed to decide control measures, implement control measures and to monitor and review measures.

The very first phase is to increase awareness about the importance of legal issues for the business within the company. There is a need to develop a strategy to bring awareness according to the roles of the employees. This should include everyone in the business. An optimum approach would be to develop training modules according to the staff hierarchy and based on their roles and responsibilities. For example marketing team should increase awareness for marketing related legal issues and IT support team should have their awareness increased about DPA, DSR, eSignatures etc. Such awareness increasing activities should be periodic.

The next phase would include introducing set of procedures to ensure legal requirements are met in various business operations such as marketing, order fulfilments, web content writing, personal information processing etc. With the assistance of legal experts senior management should create standard procedures to ensure compliance. All the checklists should be completed before completing any business operation. Use of Information Systems can effectively improve legal compliance requirements and many operations can be automated that would save time and money. IT Support Managers should ensure that the IT Systems that are being used within the company such as CRM, Marketing Systems, Accounting Systems, and Knowledge Management System etc. are compliant. Monitoring, accountability and transparency can be effectively managed through the use of legal compliance management systems. There are many commercial solutions available for such purposes.

Once the business has adapted the culture to increase awareness for legal issues, procedures are put in place and technology assistance is gained in order to comply and monitor compliance an Employee Reward System should be introduced in the organization rewarding those who exceeds the minimum legal requirements and keeps the business protected from potential legal risks. The rewards may be given in the shapes of bonuses, presents; points based systems etc. and most importantly giving recognition and appreciation for the psychological boosts of the employees. The following figure gives an overview of the approach.

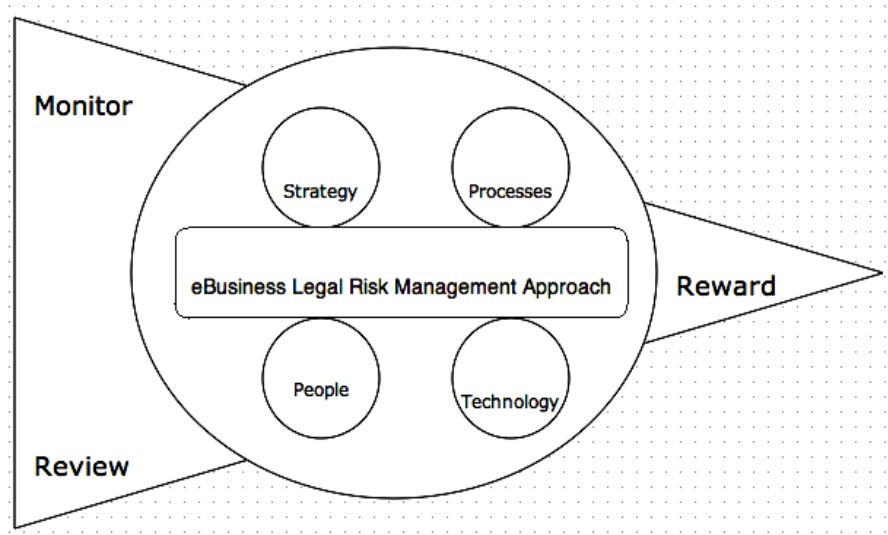

Figure 2: An approach to promote Legal Risk Management culture.

## 6. RECOMMENDATIONS TO IMPROVE INTERNATIONAL HARMONIZATION OF IT AND INTERNET LAWS

Law making process is tedious and time consuming. Technological advancements are often rapid compared to law making. However law-making authorities at national and international level should proactively assess current needs of new regulations in the field of IT. European Union Commission has made good efforts in harmonizing laws concerning IT and eBusinesses within its member countries. Several EU Directives have been discussed in previous sections, which are implemented in all member countries. This protects businesses and consumers at the same time. As in all legal areas, the area of IT and Internet is evolving gradually in line with the era's needs and will continue to evolve. However special attention should be given to important upcoming issues such as: easing marketing regulations for SMEs, making legit use of data permissible for market research purposes etc. With the ever-increasing social networking online platforms, issues of privacy and use of blogs for marketing purposes needs attention.

International harmonization of laws is the biggest challenge at the moment. Although EU has harmonized certain laws for its member countries, other countries should encourage signing treaties to bind by such regulations. Use of Internet is increasing rapidly in Asia and Africa as well and it will be global benefit to harmonies laws and cross border trade will continue to flourish as a result of globalization through telecommunication.

## 7. RECOMMENDATIONS FOR SMEs TO RUN eBUSINESSES LEGALLY

Some of the recommendations for SMEs to run eBusinesses legally and ethically are as follows:
- It should be ensured that all the contents on the Internet and all communications comply with UK law and EU Law. Then it should comply with the jurisdiction outside the UK and EU where the business is substantially targeted.

- A change to law is a continuous process, stay up to date with the relevant legislation that might affect several business processes.

- All personal data processing must be in compliance with the eight principles laid out in Data Protection Act 1998.

- Ensure that the eBusiness's Intellectual Property is protected. Making use of good IP Management System could substantially reduce the legal complexities.

- All contracts should be in written and legally binding. If electronic contracts are being used to close deals ensure that eSignature regulations have been met.

- When developing an eMarketing strategy closely consider compliance requirements and ensure that all adverts are true and honest. Any other form of marketing strategy use such as direct electronic marketing or multi-level marketing must also be compliant.

- Promote a culture in the organization where everyone contributes his or her part in ensuring organization wide legal compliance.

- Wherever needed seek expert legal advice especially while developing strategies, creating procedures and deploying Information Technology Systems.

## 8. CONCLUSIONS

In this research eBusinesses and eMarketing have been discussed from a legal perspective. Legal Risk Management is currently going through its infancy but progressing rapidly. It has become vital to have a risk management strategy in eBusinesses. In this research some of the threats to eBusinesses were identified, analysed and discussed. Later a proposed approach was overviewed to promote legal risk management culture in organizations. Providing recommendations to law making authorities to promote international harmonization of IT Laws followed this. Also recommendations for SMEs to run eBusinesses legally were listed down.

The future of IT Laws would see their gradual evolution. Latest technological advancements will create new needs for legal protection for businesses as well as consumers. Law making authorities should be proactive in assessing latest legal needs and make laws for them. The global nature of Internet is its beauty and the benefits mankind can achieve from it are countless. Cross border B2C trade has become accessible which means any local business can somehow go international. There is a big gap in progressing economies where businesses can provide their goods and services. Keeping harmonization and consistency in laws will make it a fair platform for everyone.

## 9. ACKNOWLEDGMENT

We are grateful to Allah Almighty for the courage and strength He gave us to complete the goals of this project. We are thankful for the support and encouragement given by our parents and their prayers for our success. We also obliged our Project Advisor and professors whose knowledge and guidance was paramount in the realization of my objective and keeping us motivated.